\documentclass[reprint,superscriptaddress,amsmath,amssymb,aps,prl,
nobibnotes 
]{revtex4-2}

\usepackage{graphicx}
\usepackage{dcolumn}
\usepackage{multirow} 
\usepackage[dvipsnames]{xcolor}
\usepackage{longtable}
\usepackage{booktabs}
\usepackage[T1]{fontenc}
\usepackage[utf8]{inputenc}
\usepackage{siunitx,url,mhchem,todonotes,xcolor}
\usepackage[colorlinks=true, allcolors=blue]{hyperref}

\usepackage{orcidlink}

\newcommand{\eV}{\,\si{\electronvolt}}
\newcommand{\keV}{\,\si{\kilo\electronvolt}}

\newcommand{\mysec}[1]{\textit{#1}}

\begin{document}

\title{Nuclear Charge Radius of $^9$Be from Muonic Atom Spectroscopy Using a Microcalorimeter}
\author{Ofir~Eizenberg\orcidlink{0009-0000-0489-1582}}
\affiliation{Physics Department, Technion—Israel Institute of Technology, Haifa 3200003, Israel}
\affiliation{The Helen Diller Quantum Center, Technion—Israel Institute of Technology, Haifa 3200003, Israel}

\author{Shikha Rathi\orcidlink{0000-0002-0768-5546}}
\affiliation{Physics Department, Technion—Israel Institute of Technology, Haifa 3200003, Israel}
\affiliation{Laboratoire Kastler Brossel, Sorbonne Universit\'e, CNRS, ENS-PSL Research University, Coll\`ege de France, 75005 Paris, France}

\author{Andreas~Abeln\orcidlink{0000-0002-0379-8148}}
\affiliation{Kirchhoff‐Institut f\"ur Physik, Universit\"at Heidelberg, 69120 Heidelberg, Germany}

\author{Sonia~Bacca\orcidlink{0000-0002-9189-9458}}
\affiliation{Institut f\"ur Kernphysik, Johannes Gutenberg-Universit\"at Mainz, 55128 Mainz, Germany}
\affiliation{PRISMA$^{++}$ Cluster of Excellence, Johannes Gutenberg-Universit\"at Mainz, 55128 Mainz, Germany}

\author{Gonçalo~Baptista\orcidlink{0000-0002-0375-3433}}
\affiliation{Laboratoire Kastler Brossel, Sorbonne Universit\'e, CNRS, ENS-PSL Research University, Coll\`ege de France, 75005 Paris, France}

\author{Nir~Barnea\orcidlink{0000-0001-8036-3052}}
\affiliation{The Racah Institute of Physics, The Hebrew University, Jerusalem 9190401, Israel}

\author{Noam~Burger\orcidlink{0009-0009-6347-5672}}
\affiliation{Physics Department, Technion—Israel Institute of Technology, Haifa 3200003, Israel}
\affiliation{The Helen Diller Quantum Center, Technion—Israel Institute of Technology, Haifa 3200003, Israel}

\author{Thomas~Elias~Cocolios\orcidlink{0000-0002-0456-7878}}
\affiliation{KU Leuven, Instituut voor Kern- en Stralingsfysica, 3001 Leuven, Belgium}

\author{Marie~Deseyn\orcidlink{0000-0002-1643-4079}}
\affiliation{KU Leuven, Instituut voor Kern- en Stralingsfysica, 3001 Leuven, Belgium}

\author{Tim~Egert\orcidlink{0009-0007-3706-703X}}
\affiliation{Institut f\"ur Kernphysik, Johannes Gutenberg-Universit\"at Mainz, 55128 Mainz, Germany}
\affiliation{PRISMA$^{++}$ Cluster of Excellence, Johannes Gutenberg-Universit\"at Mainz, 55128 Mainz, Germany}

\author{Christian~Enss\orcidlink{0009-0004-2330-6982}}
\affiliation{Kirchhoff‐Institut f\"ur Physik, Universit\"at Heidelberg, 69120 Heidelberg, Germany}

\author{Andreas~Fleischmann\orcidlink{0009-0007-3706-703X0000-0002-0218-5059}}
\affiliation{Kirchhoff‐Institut f\"ur Physik, Universit\"at Heidelberg, 69120 Heidelberg, Germany}

\author{Loredana~Gastaldo\orcidlink{0000-0002-7504-1849}}
\affiliation{Kirchhoff‐Institut f\"ur Physik, Universit\"at Heidelberg, 69120 Heidelberg, Germany}

\author{C\'esar~Godinho\orcidlink{0000-0003-3195-0394}}
\affiliation{Laboratoire Kastler Brossel, Sorbonne Universit\'e, CNRS, ENS-PSL Research University, Coll\`ege de France, 75005 Paris, France}
\affiliation{LIBPhys, LA-REAL, NOVA FCT, Universidade NOVA de Lisboa, 2829-516 Caparica, Portugal}

\author{Nitzan~Goldberg}
\affiliation{The Racah Institute of Physics, The Hebrew University, Jerusalem 9190401, Israel}

\author{Michael~Heines\orcidlink{0000-0001-8723-5408}}
\affiliation{KU Leuven, Instituut voor Kern- en Stralingsfysica, 3001 Leuven, Belgium}

\author{Daniel~Hengstler\orcidlink{0009-0001-0684-8067}}
\affiliation{Kirchhoff‐Institut f\"ur Physik, Universit\"at Heidelberg, 69120 Heidelberg, Germany}

\author{Paul~Indelicato\orcidlink{0000-0003-4668-8958}}
\affiliation{Laboratoire Kastler Brossel, Sorbonne Universit\'e, CNRS, ENS-PSL Research University, Coll\`ege de France, 75005 Paris, France}

\author{Weiguang~Jiang\orcidlink{0000-0001-8441-972}} 
\affiliation{Institut f\"ur Kernphysik, Johannes Gutenberg-Universit\"at Mainz, 55128 Mainz, Germany}
\affiliation{PRISMA$^{++}$ Cluster of Excellence, Johannes Gutenberg-Universit\"at Mainz, 55128 Mainz, Germany}

\author{Klaus~Kirch\orcidlink{0000-0002-1720-7636}}
\affiliation{PSI Center for Neutron and Muon Sciences, 5232 Villigen, Switzerland}
\affiliation{ETH Z\"urich, Institute for Particle Physics and Astrophysics, 8093 Z\"urich, Switzerland}

\author{Andreas~Knecht\orcidlink{0000-0002-3767-950X}}
\affiliation{PSI Center for Neutron and Muon Sciences, 5232 Villigen, Switzerland}

\author{Daniel~Kreuzberger\orcidlink{0009-0004-9291-4148}}
\affiliation{Kirchhoff‐Institut f\"ur Physik, Universit\"at Heidelberg, 69120 Heidelberg, Germany}

\author{Jorge~Machado\orcidlink{0000-0002-0383-4882}}
\affiliation{LIBPhys, LA-REAL, NOVA FCT, Universidade NOVA de Lisboa, 2829-516 Caparica, Portugal}

\author{Ulf-G.~Mei{\ss}ner\orcidlink{0000-0003-1254-442X}}
\affiliation{Helmholtz-Institut~f\"{u}r~Strahlen-~und~Kernphysik,~Cluster~of~Excellence~--~Color~meets~Flavor and
Bethe~Center~for~Theoretical~Physics, Universit\"{a}t~Bonn,~53115~Bonn,~Germany} 
\affiliation{Institute for Advanced Simulation (IAS-4), Forschungszentrum J\"{u}lich, 52425 J\"{u}lich, Germany}
\affiliation{Peng Huanwu Collaborative Center for Research and Education, International Institute for Interdisciplinary and Frontiers, Beihang University, Beijing 100191, China}

\author{Ben~Ohayon\orcidlink{0000-0003-0045-5534}}
\email[Corresponding author: ]{bohayon@technion.ac.il}
\affiliation{Physics Department, Technion—Israel Institute of Technology, Haifa 3200003, Israel}
\affiliation{The Helen Diller Quantum Center, Technion—Israel Institute of Technology, Haifa 3200003, Israel}

\author{Nancy~Paul\orcidlink{0000-0003-4469-780X}}
\affiliation{Laboratoire Kastler Brossel, Sorbonne Universit\'e, CNRS, ENS-PSL Research University, Coll\`ege de France, 75005 Paris, France}

\author{Randolf~Pohl\orcidlink{0000-0001-6435-659X}}
\affiliation{Institut f\"ur Physik, QUANTUM, Johannes Gutenberg-Universit\"at Mainz, 55128 Mainz, Germany}
\affiliation{PRISMA$^{++}$ Cluster of Excellence, Johannes Gutenberg-Universit\"at Mainz, 55128 Mainz, Germany}

\author{Tim~Redelbach\orcidlink{0009-0005-3543-1542}}
\affiliation{Institut f\"ur Physik, QUANTUM, Johannes Gutenberg-Universit\"at Mainz, 55128 Mainz, Germany}

\author{Michael~Roosa\orcidlink{0000-0002-3449-4665}}
\affiliation{Laboratoire Kastler Brossel, Sorbonne Universit\'e, CNRS, ENS-PSL Research University, Coll\`ege de France, 75005 Paris, France}

\author{Katharina~von~Schoeler\orcidlink{0009-0005-2671-5869}}
\affiliation{ETH Z\"urich, Institute for Particle Physics and Astrophysics, 8093 Z\"urich, Switzerland}

\author{Quentin~Senetaire\orcidlink{0009-0008-4437-2099}}
\affiliation{Laboratoire Kastler Brossel, Sorbonne Universit\'e, CNRS, ENS-PSL Research University, Coll\`ege de France, 75005 Paris, France}

\author{Shihang~Shen\orcidlink{0000-0002-8051-6466}}
\affiliation{Peng Huanwu Collaborative Center for Research and Education, International Institute for Interdisciplinary and Frontiers, Beihang University, Beijing 100191, China}

\author{Daniel~Unger\orcidlink{0009-0008-8204-9145}}
\affiliation{Kirchhoff‐Institut f\"ur Physik, Universit\"at Heidelberg, 69120 Heidelberg, Germany}

\author{Stergiani~Marina~Vogiatzi\orcidlink{0000-0002-4239-8673}}
\affiliation{Institut f\"ur Physik, QUANTUM, Johannes Gutenberg-Universit\"at Mainz, 55128 Mainz, Germany}
\affiliation{SFB1660, Johannes Gutenberg-Universit\"at Mainz, 55128 Mainz, Germany}
\affiliation{Institut f\"ur Kernphysik, Johannes Gutenberg-Universit\"at Mainz, 55128 Mainz, Germany}

\author{Johanna~Walch\orcidlink{0009-0001-7583-4080}}
\affiliation{Helmholtz Institute Jena, 07743 Jena, Germany}
\affiliation{GSI Helmholtzzentrum f\"ur Schwerionenforschung GmbH, 64291 Darmstadt, Germany}
\affiliation{Friedrich-Schiller-Universität Jena, 07743 Jena, Germany}

\author{Frederik~Wauters\orcidlink{0000-0001-9343-4251}}
\affiliation{Institut f\"ur Kernphysik, Johannes Gutenberg-Universit\"at Mainz, 55128 Mainz, Germany}
\affiliation{PRISMA$^{++}$ Cluster of Excellence, Johannes Gutenberg-Universit\"at Mainz, 55128 Mainz, Germany}

\author{Aziza~Zendour\orcidlink{0009-0008-8402-3715}}
\affiliation{PSI Center for Neutron and Muon Sciences, 5232 Villigen, Switzerland}
\affiliation{ETH Z\"urich, Institute for Particle Physics and Astrophysics, 8093 Z\"urich, Switzerland}

\date{\today}
      
\begin{abstract}
The $2p\to1s$ transition energy in muonic $^9$Be was measured using a metallic magnetic calorimeter, resulting in $E_{2p\to 1s}=33\,391.48(34)\,$eV. The result is 30 times more precise than the previous best measurement and enables the extraction of the corresponding nuclear charge radius $r_c($$^9$Be$)=2.5506(51)\,$fm. It is $2.4$ times more precise than the commonly used value based on electron scattering and differs from it by $2.3$ times the combined uncertainties. 
This measurement represents the first determination of a nuclear charge radius using muonic x-ray spectroscopy with microcalorimeters.
\end{abstract}

\maketitle

The accuracy of comparisons between high-precision measurements in simple atomic systems and their theoretical predictions is often limited by the available knowledge of the root-mean-square (RMS) nuclear charge radius, $r_c$~\cite{Friar2003, 2021-Spec, 2025-CKM, 2025-CODATA, 2026-LiLike}.
As $r_c$ cannot be computed from first principles with the required precision, it must be inferred from suitable measurements.
Spectroscopy of muonic atoms, which are exotic bound states of a negative muon and a nucleus, is a workhorse for precisely measuring $r_c$~\cite{Landolt2004}. This stems from their compact size, which approaches the nuclear scale.
However, the precision of $r_c$ for light nuclei with $Z=3$--$10$ is quite limited~\cite{Ohayon2024}, with uncertainties up to two orders of magnitude larger than those for their lighter~\cite{2024-RMP} or heavier~\cite{Ohayon2025} counterparts. This arises from two main reasons: first, the experimental challenges associated with performing muonic-atom laser spectroscopy above $Z=2$~\cite{Schmidt_2018}; and second, the limited resolution of conventional x-ray detectors compared to the contribution of finite nuclear size to their atomic binding energy~\cite{Schaller1980}.
Consequently, most charge radii of light nuclei beyond helium are derived from electron scattering~\cite{Ohayon2025}, making them not only relatively imprecise but also subject to incompletely known systematic effects~\cite{2026-WhitePaper}. A relevant example is the isotope $^9$Be, whose charge radius is commonly reported as $r_c = 2.519(12)\,\text{fm}$~\cite{jansen1972nuclear}. Nevertheless, because the experiment probed only a limited range of momentum-transfer values, the extracted radius relied on a model-dependent analysis.

In this work, we demonstrate a new approach for measuring charge radii in light nuclei. The method relies on high-precision muonic-atom x-ray spectroscopy conducted using a detector array based on Metallic Magnetic Calorimeters (MMCs) operated below 100\,mK~\cite{Fleischmann2005}.
These quantum-sensing detectors provide high resolution and quantum efficiency for photons with energies up to approximately $\,200\,$keV~\cite{Enss2000, Fleischmann2004, Pies2012, Kempf2018, 2022-GSI}.
We report the deployment of this detector at the Paul Scherrer Institute (PSI) and the measurement of the muonic Be $2p\to1s$ center-of-gravity transition energy around $33\,$keV to $0.3\,$eV precision, constituting a 30-fold improvement over the best prior measurement~\cite{Schaller1980}. We then extract the absolute nuclear charge radius using dedicated state-of-the-art calculations that encompass atomic and nuclear theory, and briefly discuss the implications of the new result.

\mysec{Deployment of MMC-based detector at PSI.}
To achieve the required accuracy, we rely on state-of-the-art single-photon quantum sensing technology. Specifically, we use a $maXs$-30 (Micro-calorimeter Array for X-ray Spectroscopy) detector optimized for measuring energies around 30\,keV~\cite{Hengstler2015, geist2020, Unger2021}, developed at the Kirchhoff Institute for Physics at Heidelberg University.
Under accelerator beamline conditions, the Full-Width at Half-Maximum (FWHM) resolution achieved is approximately $20\,$eV at $33\,$keV, which is 15 times better than that of the highest-resolution solid-state detectors~\cite{Knoll2010, clozza2026extended}. 

The detector has an $8 \times 8$ array of pixels coupled in pairs to a two-stage SQUID readout. Its active area is $4\times4$\,mm$^2$.
Each pixel is equipped with a $20\,\mu\text{m}$-thick gold absorber that efficiently stops photons up to 120\,keV (e.g., 95\% efficiency at 20 keV and 10\% at 120 keV)~\cite{unger2024mmc}.
The MMC array was mounted inside a custom sidearm of a dilution refrigerator, facing a target chamber, with $13\,$cm between the target center and the MMC absorber plane.
This resulted in a solid angle of $8\times10^{-5}\,$.
Given the quantum efficiency of nearly $60\%$ at $33\,$keV \cite{unger2024mmc}, and considering that approximately two-thirds of the pixels were fully functioning, 
the overall efficiency amounted to $3\times10^{-5}$ at $33\,$keV.

The measurement was carried out over a period of 34~hours at the $\pi E_1$ beamline of the PSI accelerator facility, which delivers a continuous beam of negative muons of adjustable momentum~\cite{Gerchow2023}.
The beam passed through a collimation system comprising a scintillator with a central aperture and subsequently entered the evacuated target chamber via a thin plastic scintillator, which served as the timing reference for the muonic X‑ray cascade~\cite{knecht2020study}.
The process from muon trigger to x-ray emission occurs within a few ns, which is substantially shorter than the time resolution of our detection system.
Inside the chamber, we placed a high-purity ($>99$\%) beryllium disk, $2.5\,$mm thick and $21.6\,$mm in diameter, made from naturally occurring beryllium, which consists entirely of $^9$Be.
To reduce self-absorption, it was mounted at a 45 degree angle relative to both the incident beam axis and the cryostat window.
At a beam momentum of $29.5$ MeV/c, the estimated muon stopping rate was $5$ kHz, based on a GEANT4 simulation, leading to a detected $2p \to 1s$ photon rate of $0.13$ Hz.

For each event in an MMC pixel, a $2.6$\,ms waveform from the analog output of the SQUID electronics was digitized using a 16-bit digitizer~\footnote{SIS3316 module from Struck Innovative Systeme GmbH, Harksheider Str. 102, 22399 Hamburg, Germany}.
The voltage nonlinearity of each digitizer channel was measured separately and used to correct all recorded waveforms.
Each pulse was fitted using amplitude scaling and time shifting of a template derived from 60\,keV photon pulses from an $^{241}$Am source that was continuously measured. 
We also placed a target made of mixed barium and lanthanum foils, sealed in epoxy, between the cryostat x-ray window and the target chamber. 
This target was irradiated with an x-ray tube to generate x-ray fluorescence (XRF) photons, which were collected simultaneously with the muonic x rays.
The reference lines and their measured rates are summarized in Table~\ref{tab:expected_lines}.
\begin{table}[tbp]
\centering
\begin{ruledtabular}
\caption{Main reference lines discussed in the text, ordered by descending energy.}\label{tab:expected_lines}  
 \begin{tabular}{rrllc}
     \textbf{Source} & \textbf{Line ~ ~} & {\textbf{Energy / eV}} & \textbf{Ref.} & \textbf{Rate / Hz} \\
     \hline
     \noalign{\vspace{2pt}}
     \llap{$^{241}$}Am ~ & \llap{$^{237}$}Np:\,$\gamma_{(2,0)}$ & 59540.2 (1) &\cite{Rodrigues2026}     & 0.12    \\
     XRF    ~ & La:\,$K_{\alpha_1}$\vphantom{$^{237}$}  & 33442.12(27)&\cite{nist_xraytrans}& 0.18    \\
     XRF    ~ & La:\,$K_{\alpha_2}$\vphantom{$^{237}$}  & 33034.38(26)&\cite{nist_xraytrans}& 0.11    \\
     \llap{$^{241}$}Am ~ & \llap{$^{237}$}Np:\,$\gamma_{(1,0)}$ & 33196.3(3)&\cite{Basunia2006}        & ~\,0.002 \\
     XRF    ~ & Ba:\,$K_{\alpha_1}$\vphantom{$^{237}$}  & 32193.26(7)&\cite{nist_xraytrans} & 0.28    \\
     XRF    ~ & Ba:\,$K_{\alpha_2}$\vphantom{$^{237}$}  & 31816.62(6)&\cite{nist_xraytrans} & 0.15    \\
     \llap{$^{241}$}Am ~ & \llap{$^{237}$}Np:\,$\gamma_{(2,1)}$ & 26345.5(1)&\cite{Rodrigues2026}      & 0.03    \\
     \hline
     \noalign{\vspace{2pt}}
     \textbf{Total}           &               &                        && \textbf{1.8} \\
    \end{tabular}
\end{ruledtabular}
\end{table}

\begin{figure}[tbp]
\centering
\includegraphics[width=0.49\textwidth, trim=9.5mm 9.3mm 7.5mm 7mm, clip]{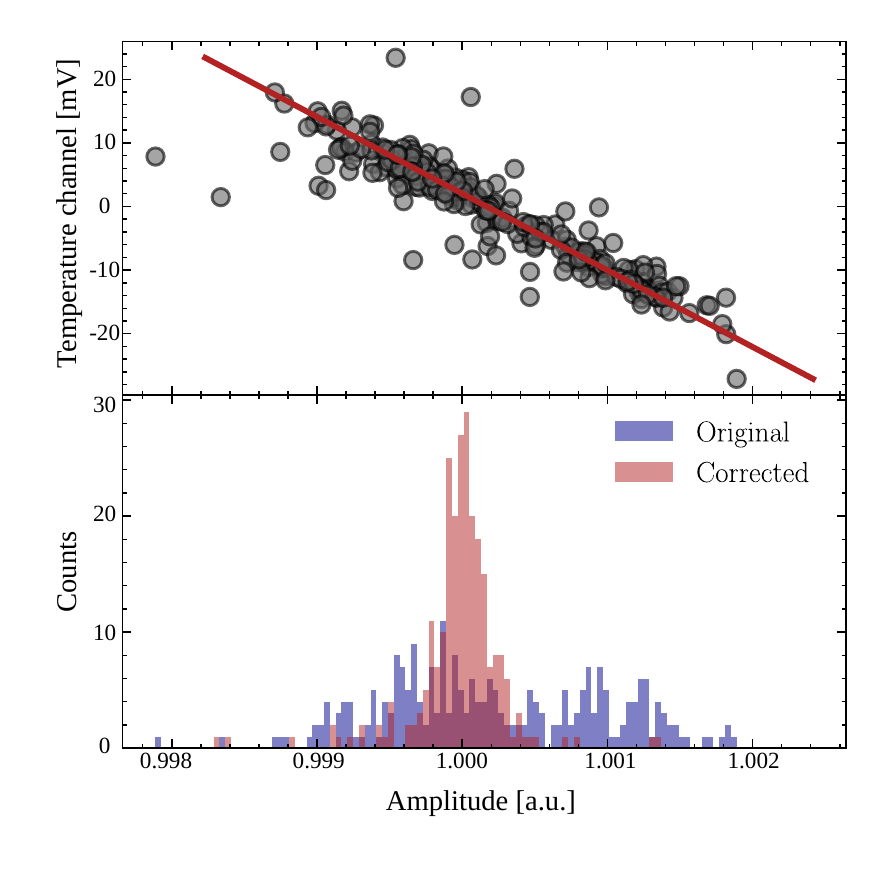}
\caption{Temperature correction for a single pixel. 
The top graph shows the correlation between the amplitude from pulses stemming from 60\,keV photons, and the baseline voltage of a temperature-sensitive channel (gray dots), along with a robust linear regression fit (red line).
The bottom graph shows how applying this linear correction drastically improves the resolution.}
\label{fig:temperature_correction}
\end{figure}

\mysec{From pulse amplitudes to summed spectrum.} 
Data reduction is applied to filter out triggered noise, deformed traces (e.g., pile-up events), and coincidence events, which can be attributed to Michel electron interactions in the detector substrate. The cuts involved in this process are described in the supplemental material. 
Despite the highly linear response of MMCs, even small temperature variations (here, a few mK) induce changes in signal amplitude that must be compensated for to achieve high energy resolution. This correction is implemented by measuring baseline voltage drifts in dedicated temperature-sensitive channels located at the four corners of the detector~\cite{unger2024mmc}.
The correction is performed separately for each pixel by evaluating the dependence of amplitude on temperature for 60~keV photons from an $^{241}$Am source and fitting with a robust linear model (using the \textit{statsmodels} package~\cite{2010-Statsmodels}).
The fitted model was used to project the entire measured spectrum to a constant temperature. 
An example of this process for one of the pixels is shown in Fig.~\ref{fig:temperature_correction}.
After correcting for the temperature dependence of the amplitude, the individual pixels and the sum of their spectra exhibit a resolving power of approximately $1700$ in the region of interest, which, after gain calibration (see below), translates to $20\,$eV\,FWHM.

\begin{figure*}[tbp]
\centering
\includegraphics[width=0.98\textwidth, trim=8mm 6mm 7.5mm 6.8mm, clip]{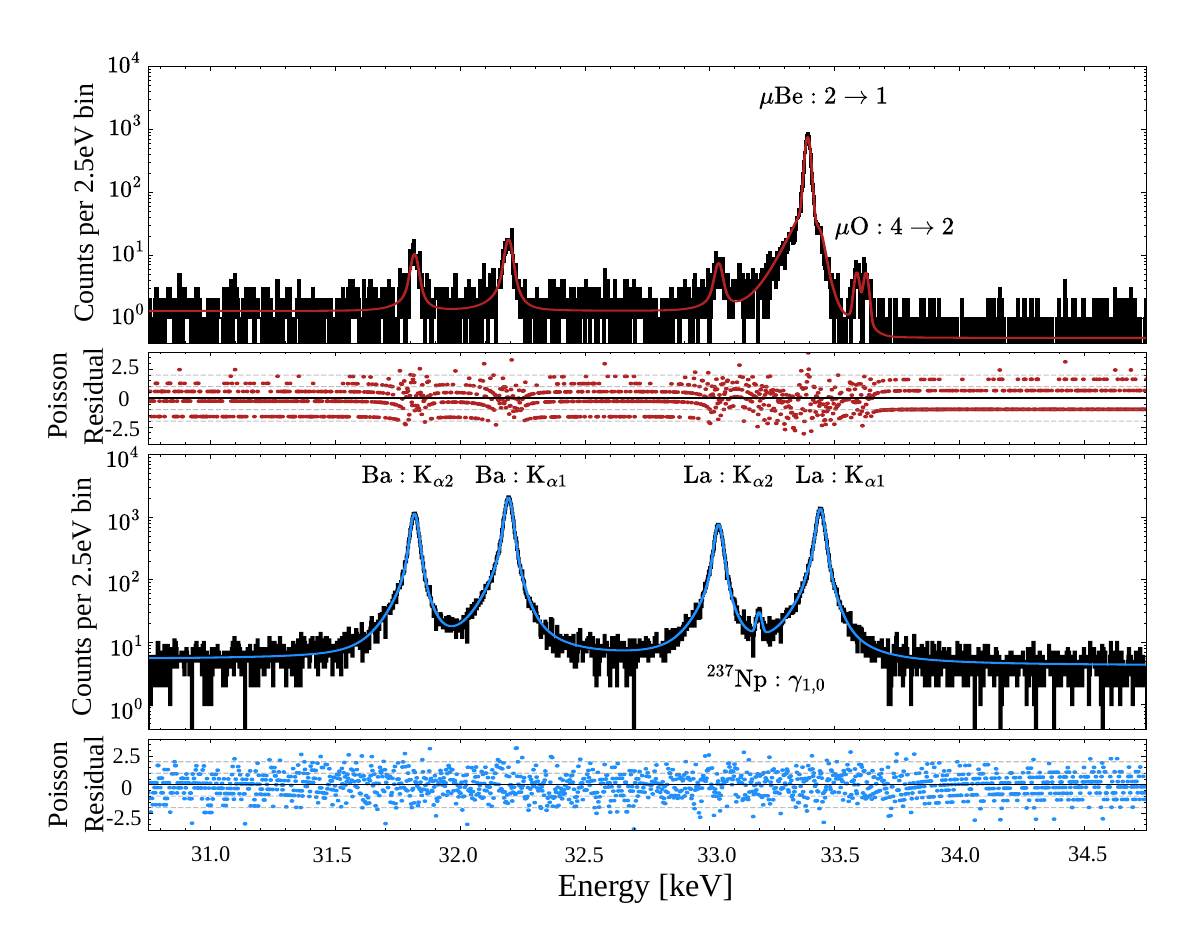}
\caption{
Fitted histograms of the combined spectra from 35 pixels in the area of interest. The top figure shows x rays in time coincidence with an incoming muon.
The bottom figure shows the anti-coincident spectrum featuring calibration lines.
}
\label{fig:fittings}
\end{figure*}

The subsequent step in the analysis involves distinguishing muon-induced events from all other events. This is crucial because our primary calibration line, La:\,$K_{\alpha_1}$, is located only about $50\,$eV away from the $\mu^9$Be:\,$2p\to1s$ transition line of interest. Partial spectral overlap can introduce lineshape distortions that hinder an unambiguous determination of the respective line centers.

To minimize bleeding of the calibration lines to the muonic x-ray spectrum, two separate spectra are constructed: one in coincidence with an incoming muon and one in anti-coincidence with it.
For each pixel, the optimal time windows relative to the muon trigger were established using muonic Be transitions that were well separated from the calibration lines, namely the $n=3\to n=2$ manifold near 6\,keV and the $3p\to1s$ transition around 39\,keV.
The chosen coincidence window is set to $750\,$ns around the most probable timing value. It reduces the calibration peak magnitude in the coincident spectrum by a factor of $125\,$ (See Fig.~\ref{fig:fittings}), without meaningfully affecting the muon-induced lines.
The remaining bleeding of the calibration signal into the muon-induced signal, and vice versa, is addressed by a coupled fitting procedure discussed below.

Although MMCs exhibit a highly linear thermodynamic response, achieving sub-eV precision necessitates a pixel-by-pixel determination of the residual nonlinearity~\cite{Sikorsky2020}.
Here, the residual non-linearity of each pixel is quantified by fitting several calibration lines (given in Table~\ref{tab:expected_lines}) located on either side of the region of interest and determining their most-probable amplitudes $A_i$. These amplitudes are then fitted according to
\begin{equation}
 E(A_i) = m(A_i+n\,A_i^2),
 \label{eq:nonlinearity}
\end{equation}
where $E(A_i)$ are the literature values of the energies given in Table~\ref{tab:expected_lines}, $m$ is a linear gain parameter, and $n$ is the pixel nonlinearity parameter. 
The fit returns $n$ between $1.65~\times$  $10^{-4}$ and $1.95~ \times$  $10^{-4}\,/\keV$, depending on the pixel, with a relative uncertainty on the order of 1\%.
The median nonlinearity around $33\,$keV is $0.6\%$, which introduces a correction of $0.3\,$eV to the approximately $50\,$eV distance between the $\mu^9$Be:\,$2p\to1s$ centroid energy and the La:\,$K_{\alpha_1}$ calibration line.
Thus, the nonlinearity correction is on the same order as our statistical uncertainty and therefore cannot be ignored. Nonetheless, it is small enough that the uncertainty associated with this correction can be neglected.
Eq.~(\ref{eq:nonlinearity}) is then used to convert pulse amplitudes to energies, and the data are summed to produce the high-statistics spectrum in Fig.~\ref{fig:fittings}.

\mysec{Final fit and energy extraction}
The energy difference between the $\mu^9$Be:\,$2p\to1s$ centroid energy and the La:\,$K_{\alpha1}$ calibration line was derived by simultaneously fitting the coincidence and anti-coincidence spectra shown in Fig.~\ref{fig:fittings}. 
The instrumental lineshape is, in the first approximation, Gaussian. However, for high precision experiments, small deviations, such as low energy tails, need to be taken into account to determine the line centroid.  Depending on the origin of the tail, an energy-dependent lineshape might be observed.  
Here, we apply a generic profile composed of a Voigt function with exponential tails. A detailed description of the lineshape 
is provided in the End Matter.

Both parts of the separated spectrum were fitted simultaneously using maximum likelihood estimation based on Poisson statistics (via the SATLAS-2 package~\cite{Gins2024}). Two additional free parameters were introduced to account for the residual cross-contamination between the spectra.
The background is modeled as constant throughout the fitting domain.
Muonic peaks exhibit characteristic step-like behavior at low energies, with the magnitude treated as an adjustable parameter in the fitting procedure. We attribute this feature to photons undergoing Compton scattering between the point of muonic-atom formation and the detector. Such effects are considerably smaller for the calibration lines because they traverse significantly less material.

The fitting model was initialized with 50 free parameters and progressively constrained based on the underlying physics, as detailed in the End Matter.
The variation of the final result induced by empirical line-shape modeling was found to be on the order of $50\,$meV; substantially smaller than the statistical uncertainty. 
The final fit has 21 free parameters, and its results are shown in Fig.~\ref{fig:fittings}.
The main outcome of the fit is the difference between the centroids of the lanthanum $K_{\alpha_1}$ and muonic $^9$Be $2p\to1s$ lines, which is
\begin{equation}\label{Eq:diff}
(\text{La}:\,K_{\alpha_1})-(\mu^9\text{Be}:\,2p\to1s)=50.71(21)[5]\,\text{eV}. 
\end{equation}
The first uncertainty in Eq.~\ref{Eq:diff} is statistical, and the second is systematic, evaluated from the difference between the results when allowing more free parameters in the fit (see End matter for more details).

The $2p\to1s$ centroid photon energy is determined by subtracting the measured energy difference from the tabulated calibration line energy (Table~\ref{tab:expected_lines}).
To convert this photon energy into the corresponding transition energy, we subtract the $66\,$meV taken by the recoiling atom, as determined from momentum conservation~\cite{1960-Lipkin}.
The final result is therefore:
\begin{equation}\label{eq:result}
     E(\mu^9\text{Be}:\,2p\to1s)= 33\,391.48(22)_\text{fit}(27)_\text{sys} \eV.
\end{equation}
Its uncertainty is primarily limited by the precision of the reference calibration line used and secondarily by the statistical precision of the experiment.

\mysec{Radius extraction and implications}
Equating our experimental result, Eq.~(\ref{eq:result}), with the theoretical parameterization detailed in the End Matter, Eq.~(\ref{eq:parC}), and solving for $r_c$ yields
\begin{equation}\label{eq:r}
     r_c(^9\text{Be}) = 2.5506(52)\,\text{fm}.
\end{equation}

Our result from Eq.~(\ref{eq:r}) is $2.4$ times more precise and larger by $2.3$ combined standard errors than the accepted value from electron scattering, $2.519(12)\,$fm~\cite{jansen1972nuclear}. 
Its total variance primarily stems from the literature value of the $K_{\alpha1}$ characteristic x ray of lanthanum~\cite{nist_xraytrans} (51\%), the fit precision of our experiment (32\%), and the uncertainty in nuclear theory for the energy to radius parametrization (17\%). See End Matter for more details.

\begin{table}[tbp]
    \centering
    \begin{ruledtabular}
    \begin{tabular}{r l l c l}
    \multicolumn{1}{c}{\textbf{Isotope}}
    & \multicolumn{1}{c}{\textbf{$r_c$(fm)}}& \multicolumn{1}{c}{\textbf{Literature}} & \multicolumn{1}{c}{\textbf{NLEFT}} & \multicolumn{1}{c}{\textbf{GFMC}}\\
\hline
\noalign{\vspace{2pt}}
    $^7$Be    ~ ~ & 2.677(12) & 2.646(16)~\cite{2012-12Be}         & 2.581(17)\\
    $^9$Be    ~ ~ & 2.5506(52)& 2.519(12)~\cite{jansen1972nuclear} & 2.554(15) &2.53(5)\\
    $^{10}$Be ~ ~ & 2.395(12) & 2.361(14)~\cite{2012-12Be}         & 2.512(36) & \\
    $^{11}$Be ~ ~ & 2.499(10) & 2.466(15)~\cite{2012-12Be}         & 2.544(40)\\
    $^{12}$Be ~ ~ & 2.535(11) & 2.502(12)~\cite{2012-12Be}         & 2.580(33)\\
    \noalign{\vspace{2pt}}
        $^{10}$C ~\,~ ~ & \textit{2.671(14)} & \textit{2.638(36)}~\cite{Ohayon2025} &&
        \\
     $^{10}$B$^*$ ~ ~ & \textit{2.562(13)} & \textit{2.531(38)}~\cite{Ohayon2025}&&
     \\
    \end{tabular}
    \end{ruledtabular}
    \caption{Updated radii derived from this work. $r_c($$^{10}$C) is derived from the mirror fit~\cite{Ohayon2025}, and $r_c($$^{10}$B$^*$), which refers to the $0^+$ excited state, is interpolated assuming negligible isospin symmetry breaking~\cite{2023-Seng, Ohayon2025}. They are presented in italics to highlight their dependence on assumptions.
    NLEFT stands for nuclear lattice effective field theory~\cite{Lahde:2019npb,Elhatisari:2022zrb,Shen:2024qzi}.
    GFMC stands for Green's function Monte Carlo~\cite{king2026quantum}. 
    }
    \label{tab:chain}
\end{table}

The radii of radioactive Be isotopes are measured using laser spectroscopy relative to the reference value~\cite{2012-12Be} and must be shifted by about $0.03\,$fm, as listed in Table~\ref{tab:chain}. This systematic shift is about 2--3 times larger than the quoted experimental uncertainties.
These radii are now better aligned with recent nuclear lattice effective field theory calculations~\cite{Shen:2024qzi}, except for $^7$Be, whose radius deviates by $4.5$ combined standard errors.
Both the literature value and our updated radius are consistent with Green's function Monte-Carlo calculations~\cite{king2026quantum} within uncertainty.

Assuming that the mirror shift relation (detailed in Ref.~\cite{Ohayon2025}) holds for light nuclei and high asymmetries, this provides an improved prediction: $r_c(^{10}$C$)=2.671(14)\,$fm. 
Further assuming negligible isospin symmetry breaking, the predicted radius of the excited $0^+$ state of $^{10}$B is $r_c(^{10}\text{B}^*)=2.565(13)\,$fm. 
These radii are important for benchmarking \textit{ab initio} calculations in superallowed beta decay~\cite{king2025quantum, 2026-deltaC, 2026-deltaC} and for estimating the magnitude and uncertainty of finite-size contributions to them~\cite{2024-ft, 2025-CKM, he2026taming}.

\mysec{Summary and outlook}
We have demonstrated the use of a detector system based on a metallic magnetic calorimeter in a negative muon beamline and performed high resolution x-ray spectroscopy of muonic atoms. The $2p\to1s$ transition energy centroid in muonic Beryllium was determined to be $33\,391.48(35)$\,eV.
From this energy and dedicated calculations spanning atomic and nuclear theory, the nuclear charge radius of $^9$Be was determined to be $2.5506(52)$\,fm
Our result is primarily limited by the literature accuracy of the calibration line used and highlights the need for improved determinations of reference line energies~\cite{Rodrigues2026}.

Combined with differential radii from optical spectroscopy~\cite{2012-12Be}, we derived improved radii for the entire chain. They are in slightly better agreement with recent \textit{ab initio} calculations, except for $^7$Be, which shows significant deviation. This radius is particularly interesting in the context of ongoing work to measure the radius of $^8$B and to derive the size of the proton halo from the difference between the two isotones~\cite{maass2020laser}.
The radii of the entire chain are now limited by the precision of measured isotope shifts, highlighting the need for new optical measurements.

In this study, we measured a line center to one percent of its linewidth, yielding an accuracy of 10 ppm. Considering that the resolution could be improved by up to a factor of 5 with an optimized detector~\cite{Unger2025}, and that x-ray line-centers have been determined to better than 500 times their linewidths~\cite{1997-Holzer,Mendenhall_2017,Mendenhall_2019}, measurements with sub ppm accuracy are within reach. 

\vspace{2 mm}
\mysec{Acknowledgments}
The experiments were performed at the $\pi E1$ beamline of PSI. We thank the accelerator and support groups for their excellent conditions. 
This research was funded in whole or in part by the Swiss National Science Foundation (SNSF) 200020\_215164.
The Technion group was supported by the Hellen Diller Quantum Center.
S. Rathi acknowledges the Technion and the Lady Davis Postdoctoral fellowships.
M. Deseyn was supported by a personal FWO Aspirant Fundamental Research (no. {11P6V24N}). 
KU Leuven BOF under contract number C14/22/104; the European Research Council (ERC) through proposal No. 101088504 (NSHAPE).
The Mainz groups were supported by the Deutsche Forschungsgemeinschaft (DFG) through the Cluster of Excellence ``Precision Physics, Fundamental Interactions, and Structure of Matter'' PRISMA$^{++}$ EXC 2118/1 (Project ID 390831469), CRC1660 "Hadrons and Nuclei as Discovery Tools" (DFG project number 514321794) and the German State of Rhineland Palatinate.
The work of U.-G. Mei{\ss}ner  was supported in part by the European Research Council (ERC) under the European Union’s Horizon 2020 research and innovation
programme (EXOTIC, grant agreement No. 101018170), the CAS President’s International Fellowship Initiative PIFI) (Grant No. 2025PD0022), Gauss Centre for Supercomputing e.V. (www.gauss-centre.eu) for computing
time on the GCS Supercomputer JUWELS at Jülich
Supercomputing Centre (JSC) and on HoreKa by the
National High-Performance Computing Center at KIT
(NHR@KIT).
The work of S. Shen was supported by the Natural Science Foundation of China under Grant No. 12435007 and U2541242.
K. von Schoeler gratefully acknowledges the support of ETH Research Grant 22-2 ETH-023, Switzerland.
J. Machado and C. Godinho acknowledge financial support from Fundação para a Ciência e Tecnologia (FCT, Portugal), under research center Grants No. UIDB/04559/2025 (LIBPhys) and No. LA/P/0117/2020 (Associated Laboratory LA-REAL), and Contract No. 2022.11197.BD, respectively.
J. Walch was supported by "Quantum Sensing for Fundamental Physics (QS4Physics)", funded by the Innovation pool of the research field Helmholtz Matter of the Helmholtz Association.
The group at the Kirchhoff Institute for Physics is supported by the Field of Focus II Initiative at Heidelberg University funded by the ``Bundesministerium f\"{u}r Bildung und Forschung” (BMBF) and by the ``Wissenschaftsministerium Baden-W\"{u}rttemberg” in the framework of the ``Exzellenzstrategie von Bund und L\"{a}ndern”.
DU acknowledges
the support of the Research Training Group HighRR (GRK 2058) funded by the Deutsche Forschungsgemeinschaft (DFG). 
\appendix

\bibliography{Biblo}

\clearpage
\newpage
\onecolumngrid
\section*{End Matter}
\twocolumngrid
\mysec{Lineshape model} 
\noindent 
A model for the response function is given in Eq.~(\ref{eq:hypermet_function}), where the signal amplitude $A$ is distributed into three components. The first component is a four-parameter Voigt core defined by the fractional intensity $f_\text{V}$, centroid $E_0$, Gaussian FWHM $\Gamma_G = 2\sigma\sqrt{2\ln(2)}$, and Lorentzian FWHM $\Gamma_L$. The remaining terms are two five-parameter tail functions located to the right and left, with fractional intensities $f_\text{rt / lt}$ and tail length scales $b_\text{rt / lt}$.
All components are summed with a constant background term $bkg$.
\begin{align}\label{eq:hypermet_function}
&f(E) = A \Big[  f_\text{voigt} \cdot V(E - E_0, \Gamma_G, \Gamma_L) \nonumber \\ \nonumber
&+ \frac{f_\text{rt}}{2b_\text{rt}} \exp\left( \frac{E - E_0}{b_\text{rt}} + \frac{\sigma^2}{2b_\text{rt}^2} \right)  \cdot \text{erfc} \left( \frac{E - E_0}{\sqrt{2}\sigma} + \frac{\sigma}{\sqrt{2}b_\text{rt}} \right) \\ \nonumber
&+ \frac{f_\text{lt}}{2b_\text{lt}} \exp\left( \frac{E_0 - E}{b_\text{lt}} + \frac{\sigma^2}{2b_\text{lt}^2} \right) \cdot \text{erfc} \left( \frac{E_0 - E}{\sqrt{2}\sigma} + \frac{\sigma}{\sqrt{2}b_\text{lt}} \right) \Big]\\ 
&+ bkg.
\end{align}
An additional step function, $A\cdot\frac{f_{\text{step}}}{2} \cdot \text{erfc} \left( \frac{E - E_0}{\sqrt{2}\sigma} \right)$, is included in the fit of the muonic peak to account for photon scattering on their way out of the target.

The muonic x-ray lines have a natural width of about $20\,$meV, which is negligible compared to our resolution. Therefore, $\Gamma_L$ for both the muonic x rays and the calibration gamma rays is taken to be zero. $\Gamma_L$ for the XRF sources is fixed to the known natural line width~\cite{krause1979}.

The fine and hyperfine structure splittings are modeled by convolving Eq.~(\ref{eq:hypermet_function}) with a sum of delta functions whose relative energies
are calculated using the MDFGME code~\cite{Indelicato2024, rathi2025reference} (see Table~\ref{tab:peak_positions_sorted}).
It was confirmed that reasonable variations in the calculations do not significantly affect the extracted centroid energies.

\begin{table}[tbp]
\centering
\caption{Calculated energies and intensities of hyperfine lines of $\mu^9\text{Be}:\,2p\to1s$ peak (sorted by descending intensity). Energies are written relative to the center-of-gravity of the $\mu^9\text{Be}:\,2p\to1s$ peak.}
\label{tab:peak_positions_sorted}

\sisetup{
    table-number-alignment = center,
    table-text-alignment  = center,
    retain-explicit-plus  = true,
    retain-zero-uncertainty = true,
    table-sign-mantissa,
    table-space-text-post = \%,
}

\begin{tabular}{|c|
                S[table-format=+1.3]  
                S[table-format=2.2]   
               |}

\hline
$(J_i, F_i)\to(J_f, F_f)$ & {\bfseries Relative Energy [eV]} & {\bfseries Intensity (\%)} \\
\hline
$(3/2, 3)\to(1/2, 2)$  &  2.356  & 29.17 \\
$(3/2, 2)\to(1/2, 2)$  &  1.942  & 11.50 \\
$(1/2, 2)\to(1/2, 1)$  & -4.133  & 11.50 \\
$(1/2, 1)\to(1/2, 2)$  &  0.012  & 11.47 \\
$(3/2, 1)\to(1/2, 1)$  & -1.516  & 11.46 \\
$(1/2, 2)\to(1/2, 2)$  & -0.170  &  9.34 \\
$(3/2, 2)\to(1/2, 1)$  & -2.021  &  9.33 \\
$(3/2, 0)\to(1/2, 1)$  & -1.029  &  4.17 \\
$(3/2, 1)\to(1/2, 2)$  &  2.447  &  1.03 \\
$(1/2, 1)\to(1/2, 1)$  & -3.951  &  1.03 \\
\hline
\end{tabular}
\end{table}

To assess the uncertainty in our empirical lineshape model, we compare the computed energy difference of interest across various models with those that impose stronger constraints. The results are summarized in table~\ref{tab:full_parameter_evolution}.

\mysec{Parametrization of energy with radius} 
Ref.~\cite{2026-Betheory} reports
\begin{align}\label{eq:par}
E_{2P-1S}(r_c)&=33\,392.55(2)\,\text{eV}\nonumber\\
&-14.32\,(r_c^2-2.519^2)\,\text{eV/fm}^2\nonumber\\
&+0.389\,(r_c^3-2.519^3)\,\text{eV/fm}^3+\Delta E_\text{NS}\,,
\end{align}
where the last term is a placeholder for corrections that depend on nuclear structure, namely
\begin{equation}\label{eq:NS}
  \Delta E_\text{NS}=\Delta E_\text{shape}+\Delta E_\text{nP}+\Delta E_\text{NP},
\end{equation}
that we refer to as shape correction, nucleon polarization, and nuclear polarization, respectively.

Ref.~\cite{2026-Betheory} assumed a Gaussian nuclear charge distribution; therefore, we include a correction for a more realistic charge distribution computed \textit{ab initio} using nuclear lattice effective field theory (NLEFT). It uses the pinhole algorithm, which inserts a sheet with specific spin-isospin labels at mid-times, allowing for the tracking of protons and neutrons along with their correlations~\cite{Elhatisari:2017eno}.
Integrating the resulting distribution yields a charge radius of $r_c^\text{NLEFT}=2.552(15)[28]\,$fm and a Friar radius of $r_z^\text{NLEFT}=3.855(26)[25]\,$fm, where the first uncertainty is numerical and the second is estimated by recalculating with a less advanced interaction (SU(4) instead of N3LO) and reporting half the difference. Both moments agree within uncertainty with recent Green's function Monte-Carlo calculations~\cite{king2026quantum}.
Their ratio is given by $(r_z/r_c)^\text{NLEFT}=1.509(2)[2]$, and it is more stable than the individual moments.
Our result is $0.3(2)\%$ smaller than that of a Gaussian distribution $(r_z/r_c)^\text{Gauss}\approx1.5146$.
Inserting this ratio into Eq.\,(7) of Ref.~\cite{2026-Betheory} gives the shape correction
\begin{equation}\label{eq:shape}
\Delta E_\text{shape}=
-0.07(4)\,\text{eV}-0.004(2)\,(r_c^3-2.519^3)~\text{eV/fm}^3,
\end{equation}
whose overall size is significant, while its sensitivity to the radius is small for any realistic variation around the reference value.
Moreover, due to the similarity of the calculated shape to that of a Gaussian, the shape correction to the elastic three-photon exchange is negligible.

The nucleon polarization is calculated using the formalism of~\cite{JiReview}, resulting in 
\begin{equation}\label{eq:nP}
\Delta E_\text{nP}= (52 \pm 49)~\text{meV}.
\end{equation}
Both the central value and its uncertainty are nearly significant.
Nuclear polarization can be assessed using an expansion 
\begin{equation}\label{eq:NP}
\Delta E_\text{NP}=\Delta E_\text{NP}^{(0)}+\Delta E_\text{NP}^{(1)}\pm\eta^2\,\Delta E_\text{NP}^{(0)}+\Delta E_\text{3pE}.
\end{equation}
The first two terms in Eq.~(\ref{eq:NP}) are the leading and next-to-leading  terms of the $\eta$ expansion of inelastic two-photon exchange~\cite{JiReview}. The third term is an uncertainty evaluated with the small parameter $\eta=\sqrt{m_\mu/m_p}\approx0.3$~\cite{JiReview,Hernandez:2019zcm}.
The last term roughly accounts for the inelastic three-photon exchange~\cite{2018-3p}.

The leading order term, $\Delta E_\text{NP}^{(0)}$, primarily depends on the electric dipole strength function, which can be predicted using a novel artificial neural network  approach~\cite{Egert2026}. For $^9\mathrm{Be}$, the prominent low-energy dipole strength-structure below 5 MeV is treated separately using a dedicated fit to the experimental strength data~\cite{Uts2000, Uts2015, Arnold2012} in this region, while the artificial neural network prediction is retained at higher energies. 
Integrating the electric dipole strength function for $^9\mathrm{Be}$ using the appropriate weight factors~\cite{JiReview} returns 
\begin{equation}\label{eq:NP0}
\Delta E_\text{NP}^{(0)}=923^{+74}_{-58}\,\text{meV},
\end{equation}
with the reported uncertainty being statistical and nucleus-dependent~\cite{Egert2026}. The magnetic contribution is negligible.

The next-to-leading term in Eq.~(\ref{eq:NP}), $\Delta E_\text{NP}^{(1)}$, is calculated using Eqs.~\,(5b)\,,(5c)\,,(5k)\, and (5l) from Ref.~\cite{JiReview}. The 1 and 2-body densities that enter these equations are taken from Ref.~\cite{Wiringa}. The process is validated against Ref.~\cite{JiReview} for helium isotopes and applied to muonic $^9$Be for densities calculated with various interaction Hamiltonians. More details will be provided in a separate publication.

Considering six Hamiltonians: AV18+UX, NV2+3-IIa*, NV2+3-IIb, NV2+3-Ia, NV2+3-Ia*, and NV2+3-Ib, the values of $\Delta E_\text{NP}^{(1)}$ range from $-45$\,meV to $23$\,meV. The full range
\begin{equation}\label{eq:NP1}
\Delta E_\text{NP}^{(1)}=-10(35)\,\text{meV},
\end{equation}
is taken as a conservative uncertainty estimate.

We follow Ref.~\cite{2024-RMP} in estimating the inelastic three photon exchange, $\Delta E_\text{3pE}$, as half of the elastic part, given by the sum of Eqs.\,(8) and\,(9) of Ref.~\cite{2026-Betheory}, with an opposite sign and 100\% uncertainty
\begin{equation}\label{eq:3pE}
\Delta E_\text{3pE}\approx 85(85)\,\text{meV}\,.
\end{equation}
Plugging Eqs.~(\ref{eq:NP0}), (\ref{eq:NP1}) and (\ref{eq:3pE}) to Eq.~(\ref{eq:NP}) results in
\begin{equation}\label{eq:NPN}
\Delta E_\text{NP}=1.00(15)\,\text{eV}.
\end{equation}
Summing Eq.~(\ref{eq:shape}), (\ref{eq:nP}), and (\ref{eq:NPN}) results in
\begin{equation}\label{eq:NSN}
\Delta E_\text{NS}=0.98(16)\,\text{eV}-0.004(2)\,(r_c^3-2.519^3)~\text{eV/fm}^3.
\end{equation}
Its uncertainty primarily arises from $\Delta E_\text{NP}^{(2)}$, followed by $\Delta E_\text{3pE}$, and subsequently from $\Delta E_\text{NP}^{(0)}$ stemming from the artificial neural network. Compared to these, the uncertainties from $\Delta E_\text{shape}$, $\Delta E_\text{NP}^{(1)}$ and the pure QED uncertainty are minor.
The complete parameterization is given by the sum of Eq.~(\ref{eq:par}) and~(\ref{eq:NSN})
\begin{align}\label{eq:parC}
E_{2P-1S}(r_c)=\,&33\,393.54(16)\,\text{eV}\nonumber\\
-14.32&\,(r_c^2-2.519^2)\,\text{eV/fm}^2\nonumber\\
+0.385(2)&\,(r_c^3-2.519^3)\,\text{eV/fm}^3\,.
\end{align}

\newcommand{\acc}[2]{\makebox[4em][r]{#1}\,/\,\makebox[4em][l]{#2}}
\newcommand{\ditto}{\textquotedbl{}}
\newcommand{\NA}{*}
\begin{table*}[t]
\centering
\caption{Complete evolution of the fit parameters across hierarchical fitting stages. Values in parentheses represent the statistical uncertainty in the last digits. "Linked" indicates the parameter was mathematically constrained to a master variable (typically the La $K_{\alpha,1}$ transition). "Fixed" indicates the parameter was locked to a literature value~\cite{krause1979} or zero. Amplitudes and step fractional intensities are denoted in both anticoincident (AC) and coincident (C) spectra, separated by `/'.}
\label{tab:full_parameter_evolution}
\begin{tabular}{l l c c c c}
\toprule
\textbf{Parameter} & \textbf{Peak} & \textbf{Unconstrained} & \textbf{Relaxed} & \textbf{Intermediate} & \textbf{Constrained} \\
 & & [50 Free] & [36 Free] & [24 Free] & [21 Free] \\
\midrule
\textbf{Test statistic} 
    & $\chi^2_{r}$ & 1.132 & 1.132  & 1.140  & 1.169 \\
\midrule

\multirow{2}{*}{\textbf{Global}} 
    & $E_{\text{diff.}}$ (eV) & 50.81(57) & 50.63(48) & 50.76(24) & 50.71(21) \\
    & Scale & 1.00001(1) & 1.00001(1) & 1.000015(7)&  1.000018(5) \\
\specialrule{.1pt}{1pt}{1pt}
\multirow{11}{*}{\begin{tabular}{@{}l@{}}\textbf{Amplitudes} \\ \textbf{(AC / C)}\end{tabular}}
    & La$:\,K_{\alpha_1}$ & \acc{1393(27)}{6(5)}  & \acc{1396(17)}{12(2)} & \acc{1405(16)}{9(1)} & \acc{1394(16)}{12(1)} \\
    & La$:\,K_{\alpha_2}$ & \acc{772(21)}{7(1)}   & \acc{768(10)}{Linked}      & \acc{782(9)}{Linked}      & \acc{780(9)}{Linked} \\
    & Ba$:\,K_{\alpha_1}$ & \acc{2140(40)}{17(2)} & \acc{2133(27)}{18(1)} & \acc{2099(21)}{18(1)} & \acc{2091(22)}{Linked} \\
    & Ba$:\,K_{\alpha_2}$ & \acc{1140(24)}{10(1)} & \acc{1151(15)}{Linked}      & \acc{1156(12)}{Linked}      & \acc{1150(12)}{Linked} \\
    & Am        & \acc{19(3)}{---}      & \acc{18(3)}{---}      & \acc{18(3)}{---}      & \acc{18(3)}{---} \\
    & Be        & \acc{$<19$}{848(20)}  & \acc{5(1)}{834(18)}  & \acc{17(7)}{791(14)}  & \acc{17(7)}{786(14)} \\
    & O$_1$         & \acc{---}{5(1)}      & \acc{---}{5(1)}       & \acc{---}{5(1)}       & \acc{---}{4(1)} \\
    & O$_2$     & \acc{---}{5(1)}       & \acc{---}{5(1)}       & \acc{---}{5(1)}       & \acc{---}{4(1)} \\

\specialrule{.1pt}{1pt}{1pt}

\multirow{3}{*}{\begin{tabular}{@{}l@{}}\textbf{Centroids} \\ \textbf{(\eV)}\end{tabular}}
    & Ba$:\,K_{\alpha_2}$ & 31817.4(5) & 31816.7(4) & 31817.1(2) & Linked \\
    & Ba$:\,K_{\alpha_1}$ & 32193.4(5) & 32193.5(5) & 32193.8(2) & Linked \\
    & La$:\,K_{\alpha_2}$ & 33034.7(8) & 33035.0(4) & Linked & Linked \\
    & Am & 33197(2) & 33197(2) & Linked & Linked \\
    & O$_1$ & 33587(3) & 33587(3) & 33588(2) & 33588(2) \\
    & O$_2$ & 33625(3) & 33625(3) & Linked & Linked \\
\specialrule{.1pt}{1pt}{1pt}
\multirow{1}{*}{\textbf{$\Gamma_L$ (\eV)}}
    & All Peaks & Fixed & Fixed & Fixed & Fixed \\
\specialrule{.1pt}{1pt}{1pt}
\multirow{5}{*}{\textbf{$\Gamma_G$ (\eV)}}
    & La$:\,K_{\alpha_1}$ & 22(1) & 22.2(4) & 22.4(3) & 23.0(3) \\
    & La$:\,K_{\alpha_2}$ & 22(1) & Linked & Linked & Linked \\
    & Ba$:\,K_{\alpha_1}$ & 22(1) & Linked & Linked & Linked \\
    & Ba$:\,K_{\alpha_2}$ & 23(1) & Linked & Linked & Linked \\
    & Be & 22(1) & Linked & Linked & Linked \\
\specialrule{.1pt}{1pt}{1pt}

\multirow{5}{*}{\textbf{$f_{\text{voigt}}$}}
    & La$:\,K_{\alpha_1}$ & 0.86(3) & 0.84(2) & 0.86(1) & 0.89(1) \\
    & La$:\,K_{\alpha_2}$ & 0.81(6) & Linked & Linked & Linked \\
    & Ba$:\,K_{\alpha_1}$ & 0.80(4) & 0.83(3) & Linked & Linked \\
    & Ba$:\,K_{\alpha_2}$ & 0.91(3) & Linked & Linked & Linked \\
    & Be & 0.88(2) & 0.88(2) & Linked & Linked \\
\specialrule{.1pt}{1pt}{1pt}

\multirow{5}{*}{\textbf{$b_{\text{rt}}$} (\eV)}
    & La$:\,K_{\alpha_1}$ & 28(4) & 26(3) & 23(1) & 25(2) \\
    & La$:\,K_{\alpha_2}$ & 24(4) & Linked & Linked & Linked \\
    & Ba$:\,K_{\alpha_1}$ & 20(2) & 21(2) & Linked & Linked \\
    & Ba$:\,K_{\alpha_2}$ & 30(9) & Linked & Linked & Linked \\
    & Be & 22(3) & 19(3) & Linked & Linked \\
\specialrule{.1pt}{1pt}{1pt}

\multirow{5}{*}{\textbf{$b_{\text{lt}}/b_{\text{rt}}$}}
    & La$:\,K_{\alpha_1}$ & 1.7(3) & 1.8(3) & 2.1(1) & 2.1(2) \\
    & La$:\,K_{\alpha_2}$ & 1.9(4) & Linked & Linked & Linked \\
    & Ba$:\,K_{\alpha_1}$ & 2.4(3) & 2.4(2) & Linked & Linked \\
    & Ba$:\,K_{\alpha_2}$ & 1.8(6) & Linked & Linked & Linked \\
    & Be & 2.5(4) & 2.9(5) & Linked & Linked \\
\specialrule{.1pt}{1pt}{1pt}

\multirow{5}{*}{\textbf{$f_{\text{lt}}/f_{\text{rt}}$}}
    & La$:\,K_{\alpha_1}$ & 1.1(5) & 1.3(4) & 1.5(2) & 1.2(2) \\
    & La$:\,K_{\alpha_2}$ & 1.6(8) & Linked & Linked & Linked \\
    & Ba$:\,K_{\alpha_1}$ & 2.1(7) & 2.0(5) & Linked & Linked \\
    & Ba$:\,K_{\alpha_2}$ & 0.9(6) & Linked & Linked & Linked \\
    & Be & 1.9(5) & 2.0(6) & Linked & Linked \\
\specialrule{.1pt}{1pt}{1pt}

\multirow{2}{*}{\begin{tabular}{@{}l@{}}\textbf{$f_{\text{step}} [\cdot 10^{-4}]$} \\ \textbf{(AC / C)}\end{tabular}}
    & Be &  9.2(9) &  9.3(9) &  10.1(8) &  10.1(8) \\
    & & & & & \\
\specialrule{.1pt}{1pt}{1pt}

\multirow{2}{*}{\begin{tabular}{@{}l@{}}\textbf{Background} \\ \textbf{(AC / C)} \end{tabular}}
    &  & 4.2(2) / 0.50(5) & 4.2(2) / 0.49(5) & 4.2(1) / 0.48(5)& 4.2(2) / 0.48(5) \\
    & & & & & \\

\bottomrule
\end{tabular}
\end{table*}
\clearpage
\newpage
\section*{Supplementary Material}

\setcounter{figure}{0}
\setcounter{equation}{0}
\setcounter{section}{0}
\setcounter{subsection}{0}

\subsection{Analysis cuts}\label{app:cuts}
\noindent
The data selection criteria are detailed in Refs.~\cite{Fleischmann2005, 2025-thesis}. To satisfy these criteria, we set specific parameters for each of the 35 instrumented pixels in this experiment.
The implementation of the selection procedure is as follows:
\begin{enumerate}
    \item Select an upper $\chi^2$ threshold that lies above the vertical dense lines corresponding to photonic peaks (see the example in Fig.~\ref{fig:chi2_cut}, top left).
    \item Exclude the initial and final segments of the run that experienced instabilities. Within the main segment, a global outlier rejection for temperature-values was applied to exclude non-physical readout errors. It was complemented by also removing deviations greater than $3\sigma$ from the local moving average as outliers. 
    See (Fig.~\ref{fig:chi2_cut}, bottom left).
    \item Omit data from pixels exhibiting unstable baselines or irregular pulse shapes. 
    \item 
    To remove events from decay electrons that heat the entire array, apply a hold-off time of $100\,$ms after each cluster event (see Fig.~\ref{fig:chi2_cut}, top right).
    \item To prevent pile-up effects, apply a hold-off time of $540\,$ms after a photon hits the same pixel (Fig.~\ref{fig:chi2_cut} (Bottom-Right)).

\end{enumerate}
The resulting rates are summarized in Table~\ref{tab:ratescuts}.
\setcounter{figure}{0}
\renewcommand{\thefigure}{S\arabic{figure}}
    \begin{figure}[hbp] 
    \centering \includegraphics[width=0.45\textwidth, trim=4.5mm 2.5mm 1mm 2.5mm, clip]{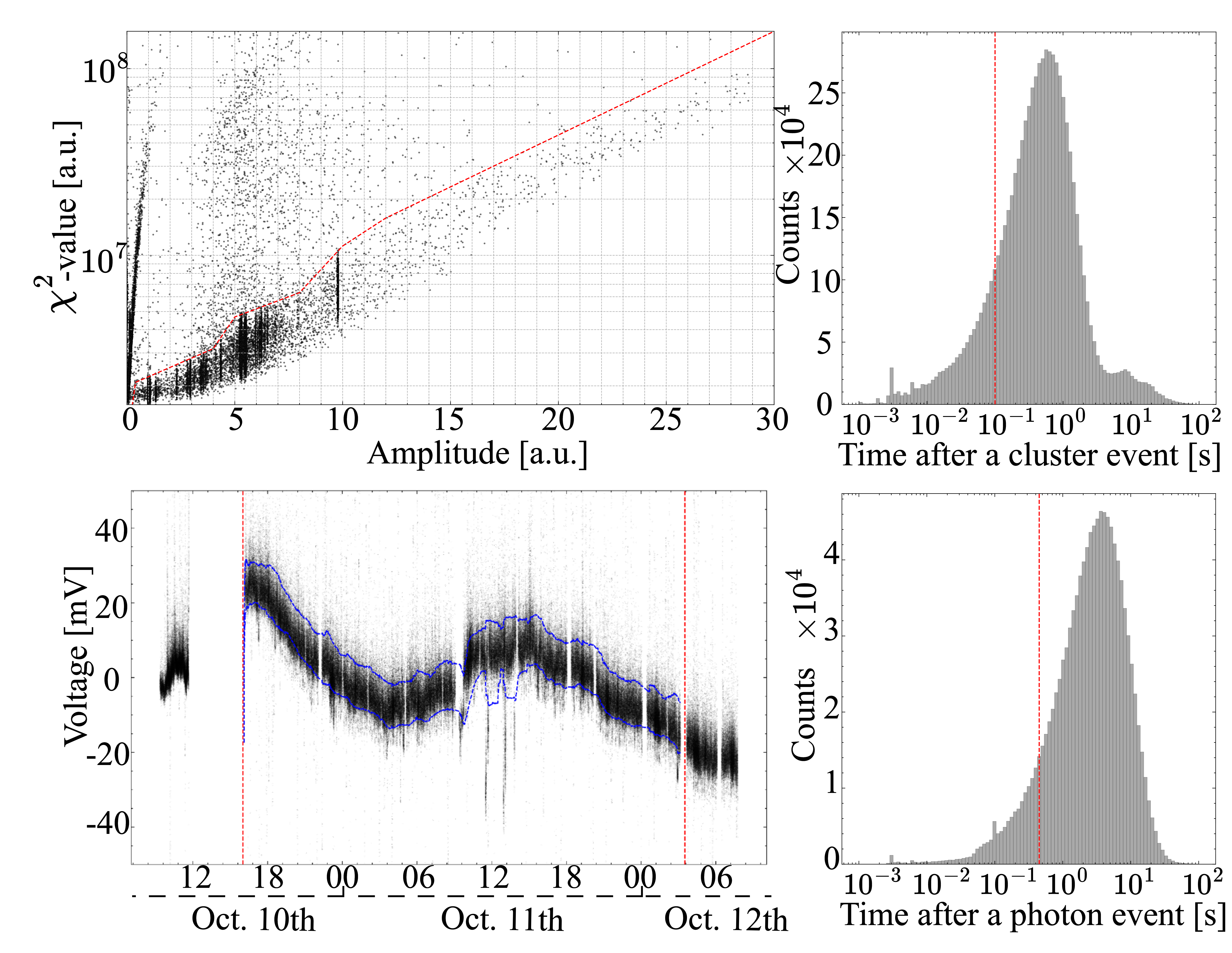 } 
    \caption{Examples of cuts described in the main text.}
    \label{fig:chi2_cut} 
    \end{figure}
\sisetup{
    table-number-alignment = center,
    table-text-alignment   = center,
    retain-explicit-plus   = true,
    retain-zero-uncertainty = true,
    table-sign-mantissa,
    table-space-text-post = \%,
}
\setcounter{table}{0}
\renewcommand{\thetable}{S\arabic{table}}
\newcounter{rownum}
\newcommand{\rownumber}{\stepcounter{rownum}\arabic{rownum}}
\begin{table}[htbp]
\centering
\setcounter{rownum}{0} 
\renewcommand{\arraystretch}{1.4} 
\caption{Summary of sequential data cuts, remaining counts, and rates summed across all active pixels of the MMC.}\label{tab:ratescuts} 
\begin{tabular}{|c|l|S[table-format=5.0]S[table-format=7.0]|S[table-format=1.2]S[table-format=1.2]|}
\hline
\multirow{2}{*}{\#} & \multirow{2}{*}{\textbf{Cut}} & \multicolumn{2}{c|}{\bfseries Counts} & \multicolumn{2}{c|}{\bfseries Rate (Hz)} \\
\cline{3-6}
& & {$_{\mu}\text{Be}$} & {La} & {$_{\mu}\text{Be}$} & {La} \\
\hline
\rownumber & $\chi^2$ cut        & 18166 & 46265 & 0.13 & 0.28 \\
\rownumber & Irregular Events    & 15548 & 35779 & 0.13 & 0.28 \\
\rownumber & Unstable Pixels     & 15138 & 34948 & 0.12 & 0.28 \\
\rownumber & Cluster Pileup      & 12818 & 29888 & 0.10 & 0.24 \\
\rownumber & Pixel Pileup        & 11741 & 27315 & 0.10 & 0.22 \\

\hline
\end{tabular}

\end{table}

\newpage
\subsection{Peak identification}\label{app:peakhunting}
\begin{figure*}[htb]
\centering
    \includegraphics[width=0.92\textwidth, trim=0 6mm 0 0, clip]{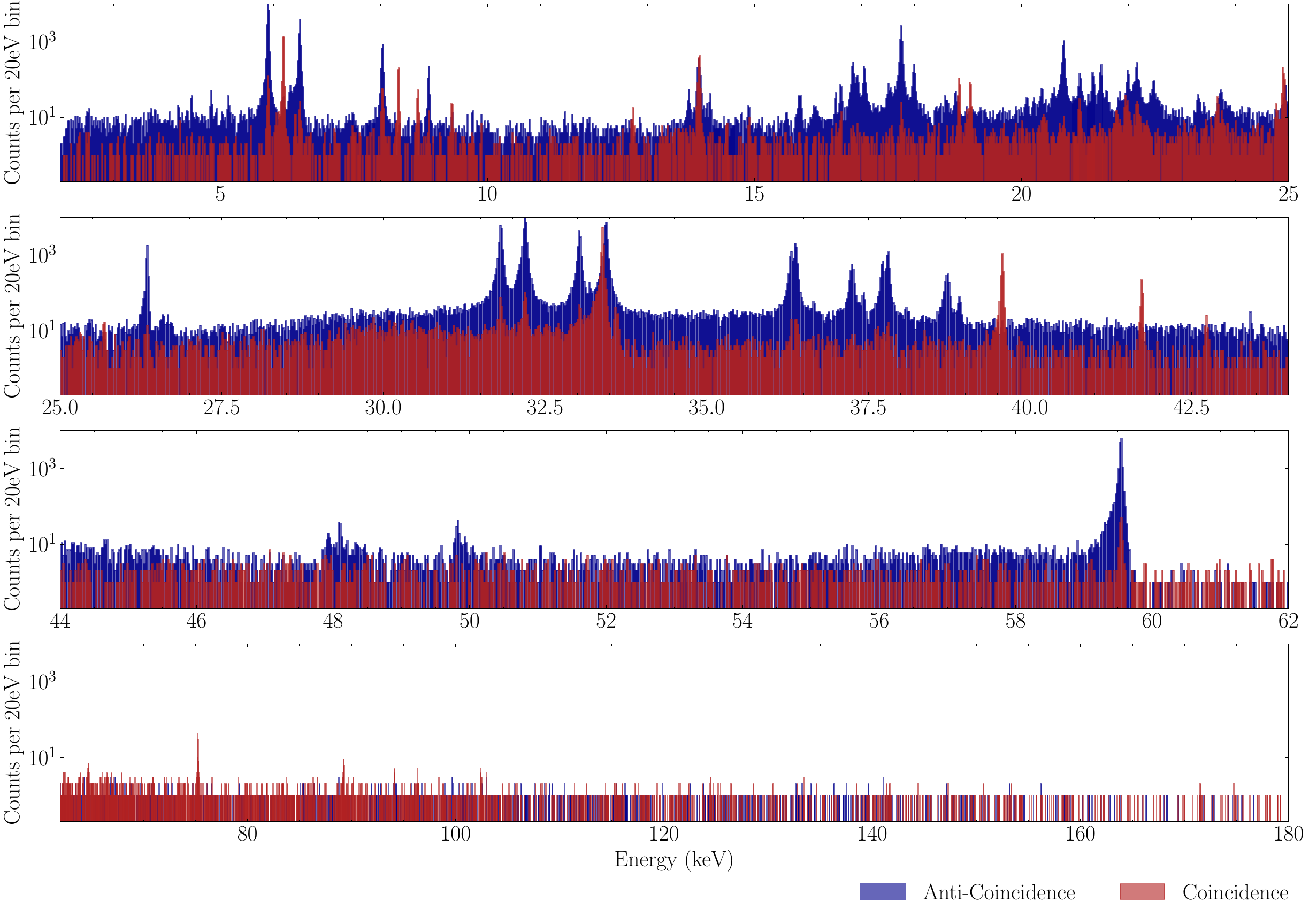}
    \caption{Coincidence and anti-coincidence spectra are shown in red and blue, respectively. The peaks are identified in Table~\ref{tab:combined_muonic_peaks}.}
    \label{fig:spectrum}
\end{figure*}

A systematic peak-hunting procedure was performed to resolve the complex, high-resolution measured spectrum. The exceptional energy resolution of the MMC enables a clear distinction between the muonic Be transitions of interest, the secondary muonic background from the target holder, and the various X-ray fluorescence and radioactive source lines used for energy scale anchoring. Table~\ref{tab:combined_muonic_peaks} and Fig.~\ref{fig:spectrum} provide an assignment of these features.

\newpage
\begin{table*}[tbph]
\footnotesize 
\renewcommand{\arraystretch}{0.9} 
\setlength{\tabcolsep}{3pt} 

\caption{List of identified peaks in the broad spectrum shown in Fig.~\ref{fig:spectrum}.  \vspace{8pt}
}
\label{tab:combined_muonic_peaks}
\begin{minipage}[t]{0.48\textwidth}
    \centering
    \textbf{Coincidence Spectrum} \\ \vspace{2pt}
    \begin{tabular}{lll}
        \toprule
        Assignment & Comments & Energy (keV) \\
        \midrule
        \multicolumn{3}{l}{\textbf{Beryllium (Target)}} \\
        $\mu \text{Be}:\,2\to1$ & Peak of Interest & 33.4\\
        $\mu \text{Be}:\,2\to1-\text{Au}:\,L_\alpha$ & Escape & 23.6\\
        $\mu \text{Be}:\,2\to1-\text{Au}:\,L_\beta$ & Escape & 22.0\\
        $\mu \text{Be}:\,3\to1$ & Trigger alignment & 39.6\\
        $\mu \text{Be}:\,4,5\to1$ & & 41.8, 42.9\\
        $\mu \text{Be}:\,3\to2$ & Trigger alignment & 6.1\\
        $\mu \text{Be}:\,4\to2$ & & 8.3\\
        \midrule
        \multicolumn{3}{l}{\textbf{Oxygen, Nitrogen, Carbon  (Target holder)}} \\
        $\mu \text{O}:\,3\to2$ & & 25.0 \\
        $\mu \text{N}:\,3\to2$ & & 19.0\\
        $\mu \text{C}:\,2\to1$ & & 75.0 \\
        $\mu \text{C}:\,3,4\to2$ & & 14.0, 18.8\\
        \midrule
        \multicolumn{3}{l}{\textbf{Copper (Chamber)}} \\
        Cu$:\,K_\alpha$, K$_\beta$ & & 8.0, 8.9 \\
        \midrule
        \multicolumn{3}{l}{\textbf{Niobium (X-Ray Shield)}} \\
        Nb$:K\alpha_{1,2}$ & & 16.5, 16.6 \\
        \bottomrule
    \end{tabular}
\end{minipage}
\hfill
\begin{minipage}[t]{0.48\textwidth}
    \centering
    \textbf{Anti-coincidence Spectrum} \\ \vspace{2pt}
    \begin{tabular}{lll}
        \toprule
        Assignment & Comments & Energy (keV) \\
        \midrule
        \multicolumn{3}{l}{\textbf{Americium-241 (Source)}} \\
        $^{237}\text{Np}:\,\gamma_{2,0}$ & Calibration & 59.5 \\
        $^{237}\text{Np}:\,\gamma_{2,0}-\text{Au}:L_{\alpha,\beta}$  & Escape & 49.8, 48.0 \\
        $^{237}\text{Np}:\,\gamma_{4,2}, \gamma_{1,0}$ & & 43.4, 33.2 \\
        $^{237}\text{Np}:\,\gamma_{2,1}$ & Calibration & 26.3 \\
        $^{237}\text{Np}:\,L_{\alpha, \beta, \gamma}$ series & & 13.8--21.0 \\
        \midrule
        \multicolumn{3}{l}{\textbf{Lanthanum (XRF Target)}} \\
        La$:\,K_{\alpha_{1,2}}$ & Calibration & 33.4, 33.0 \\
        La$:\,K_\beta, _\gamma$ series & & 37.7--39.0 \\
        La$:\,K_\alpha-\text{Au}:L_{\alpha,\beta}$  & & 21.7, 23.4, 23.8 \\
        \midrule
        \multicolumn{3}{l}{\textbf{Barium (XRF Target)}} \\
        Ba$:\,K_{\alpha_{1,2}}$ & Calibration & 32.1, 31.8 \\
        Ba$:\,K_\beta, _\gamma$ series & & 36.3--37.5 \\
        Ba$:\,K_\alpha-\text{Au}:L_{\alpha,\beta}$  & Escape & 20.3, 22.0, 22.3 \\
        Ba$:\,L_\alpha,_\beta$ series &  & 4.5, 4.8, 5.1 \\
        \midrule
        \multicolumn{3}{l}{\textbf{Iron-55 (Source)}} \\
        Mn$:\,K_\alpha, K_\beta$ &  & 5.9, 6.5\\
        \midrule
        \multicolumn{3}{l}{\textbf{Copper (Chamber)}} \\
        Cu$:\,K\alpha, K\beta$ & & 8.0, 8.9 \\
        \midrule
        \multicolumn{3}{l}{\textbf{Niobium (X-Ray Shield)}} \\
        Nb$:\,K_{\alpha_{1,2}}$ & & 16.5, 16.6 \\
        \midrule
        \multicolumn{3}{l}{\textbf{Silver (Detector)}} \\
        Ag$:\,K_{\beta}$ & & 24.9 \\
        \bottomrule
    \end{tabular}
\end{minipage}
\end{table*}

\end{document}